\documentclass[aps,pre,twocolumn,superscriptaddress,showkeys]{revtex4}
\usepackage{graphics}
\usepackage[pdftex]{graphicx}
\usepackage{times}
\usepackage{color}
\usepackage{float}
\usepackage[pdftex]{hyperref}
\hypersetup{colorlinks=false,pdfborder=0}
\usepackage{color}
\begin{document}

\title{High-performance parallel computing in the classroom using the public goods game as an example}

\author{Matja{\v z} Perc}
\thanks{Electronic address: \href{mailto:matjaz.perc@uni-mb.si}{\textcolor{blue}{matjaz.perc@uni-mb.si}}}
\affiliation{Faculty of Natural Sciences and Mathematics, University of Maribor, Koro{\v s}ka cesta 160, SI-2000 Maribor, Slovenia}
\affiliation{CAMTP -- Center for Applied Mathematics and Theoretical Physics, University of Maribor, Mladinska 3, SI-2000 Maribor, Slovenia}

\begin{abstract}
The use of computers in statistical physics is common because the sheer number of equations that describe the behavior of an entire system particle by particle often makes it impossible to solve them exactly. Monte Carlo methods form a particularly important class of numerical methods for solving problems in statistical physics. Although these methods are simple in principle, their proper use requires a good command of statistical mechanics, as well as considerable computational resources. The aim of this paper is to demonstrate how the usage of widely accessible graphics cards on personal computers can elevate the computing power in Monte Carlo simulations by orders of magnitude, thus allowing live classroom demonstration of phenomena that would otherwise be out of reach. As an example, we use the public goods game on a square lattice where two strategies compete for common resources in a social dilemma situation. We show that the second-order phase transition to an absorbing phase in the system belongs to the directed percolation universality class, and we compare the time needed to arrive at this result by means of the main processor and by means of a suitable graphics card. Parallel computing on graphics processing units has been developed actively during the last decade, to the point where today the learning curve for entry is anything but steep for those familiar with programming. The subject is thus ripe for inclusion in graduate and advanced undergraduate curricula, and we hope that this paper will facilitate this process in the realm of physics education. To that end, we provide a documented source code for an easy reproduction of presented results and for further development of Monte Carlo simulations of similar systems.
\end{abstract}


\maketitle

\section{Introduction}
The use of computers to solve problems in statistical physics has a long and fruitful history, dating as far back as the Manhattan Project, where analog computers were used so frequently they often broke down. Digital computers, such as the ENIAC (Electronic Numerical Integrator and Computer), were intertwined with nuclear science from its onset onwards. In fact, one of the first real uses of ENIAC was by Edward Teller, who used the machine in his early work on nuclear fusion reactions \cite{manhattan}. Today, computers are used in practically all areas of physics, and it is indeed difficult to imagine scientific progress without them. As rightfully pointed out by Newman and Barkema \cite{newman_99}, Monte Carlo methods form the largest and the most important class of numerical methods for solving statistical physics problems. Not surprisingly, in addition to fascinating original research dating back more than three decades \cite{binder_prb74, binder_prb80, binder1981finite}, the subject is covered in reviews \cite{binder_rpp97, hinrichsen_ap00, odor_rmp04} and many books \cite{binder_88, newman_99, marro_99, landau_00} in varying depth.

While the famous Ising model \cite{ising1925beitrag, onsager_pr44} is the workhorse behind many introductory as well as less introductory texts on the subject \cite{ibarra2016hobbyhorse, newman_99}, and is likely the most thoroughly researched model in the whole of statistical physics \cite{krauth2006statistical}, we here use another example for two reasons. In the first place, how to adapt the classical Monte Carlo algorithm for the Ising model to become fit for parallel computing on a graphics processing unit has already been demonstrated by Preis et al. \cite{preis2009gpu}. In fact, parallel computing has already been applied to a number of other statistical physics problems, such as to the $(1+1)$ dimensional surface growth model \cite{schulz2011simulation}, to the Kardar-Parisi-Zhang model \cite{kelling2012comparison, kelling2011extremely, odor2014aging}, to simulate stochastic differential equations \cite{janus2010acc}, Brownian motors \cite{spiechowicz2015gpu}, stochastic processes \cite{barros2011simulation}, and Arnold diffusion \cite{seibert2011mapping}, as well as to study ferrofluids \cite{polyakov2013large}, photon migration \cite{alerstam2008parallel}, anomalous coarsening in the disordered exclusion process \cite{juhasz2012anomalous}, and probability-based simulations in general \cite{tomov2005benchmarking}. Secondly, it might be welcome to add a little color to the curriculum by expanding on the classical subjects and thus to increase the popularity of physics with the students, although the public goods game has been a fixture in statistical physics research for at least a decade \cite{szabo_pr07, perc_jrsi13}.

The public goods game that we use here as an example is often studied in the realm of evolutionary game theory as the paradigmatic case of a social dilemma \cite{sigmund_93, weibull_95, hofbauer_98, nowak_06, sigmund_10}. The blueprint of the game is simple. The public goods game is played in groups, wherein each member of the group can choose between two strategies. If a member chooses to cooperate, it contributes a fixed amount to the common pool ($c=1$). Conversely, if a player chooses to defect, it contributes nothing to the common pool ($c=0$). The contributions from all the cooperators within a group are summed together and multiplied by a so-called synergy factor $r>1$. The latter takes into account the added value of a group effort. Lastly, the sum total of public goods after the multiplication is divided equally among all group members, and this regardless of their strategies. It is thus straightforward to see that an individual member is best of if it chooses to defect, because it can enjoy the same benefits as cooperators whilst contributing nothing. However, if everybody chooses to defect the factor $r$ multiplies zero and the public goods are lost to all. The latter scenario is often referred to as the tragedy of the commons \cite{hardin_g_s68}. Hence the classical social dilemma is given, where what is best for an individual is at odds with that is best for the group or the society as a whole. The question is under which conditions cooperation can nevertheless evolve.

Methods of statistical physics have recently been applied to subjects that, in the traditional sense, could be considered as out of scope. Statistical physics of social dynamics \cite{castellano_rmp09}, of evolutionary games in structured populations \cite{szabo_pr07, perc_bs10, wang_z_epjb15}, of crime \cite{orsogna_plr15}, and of epidemic processes and vaccination \cite{pastor_rmp15, wang_z_pr16}, are all recent examples of this exciting development. And the evolution of cooperation in the realm of the public goods game is no exception \cite{perc_jrsi13, perc_pla16}. Especially the consideration of the public goods game in a structured population is within the domain of statistical physics \cite{szolnoki_pre09c, szolnoki_prx13, javarone2016conformity, javarone2016role}. In the simplest case, a structured population is described by the square lattice, whereon cooperators can form compact clusters and can thus avoid, at least those in the interior of such clusters, being exploited by defectors \cite{nowak_n92b}.

When studying the public goods game on a square lattice, Monte Carlo simulations are used for random sequential strategy updating. This ensures that the treatment is aligned with fundamental principles of statistical mechanics, and it enables a comparison of obtained results with generalized mean-field approximations \cite{dickman_pre01, szolnoki_pre02, dickman_pre02, szolnoki_pre05} as well as a proper determination of phase transitions between different stable strategy configurations. However, such monte Carlo simulations require significant computational resources, especially if the size of the lattice is large, and if the system is close to a phase transition where fluctuations are strong. It is thus of interest to utilize parallel computing that is nowadays possible on many graphic cards installed in personal computers. Here we use the NVIDIA graphic card GeForce GTX 1080 and the CUDA programming environment \cite{cudasource}. The latter is designed to work with all main programming languages, including C that we use, as well as C++, C\#, Fortran, Python and Java. Graphic cards that support the CUDA programming environment are today available from hundred euros upward from all main graphic card manufacturers.

In what follows, we briefly describe the CUDA programming environment in Section~II, and we describe the public goods game and the parallelization of the Monte Carlo method in Sections~III and IV, respectively. In Section~V we present the results and compare the performance of parallel computing with traditional CPU-based computing. Lastly, we sum up and discuss the development of similar Monte Carlo simulations for related systems in Section~VI.

\section{CUDA programming environment}
The CUDA programming environment is freely available and upon installation integrates itself seamlessly with existing compilers available in the operating system. It includes the CUDA compiler, math libraries, as well as tools for debugging and performance optimization. The documentation available online at \cite{cudasource} is comprehensive and clear, so the reader is advised to look for details there.

The CUDA extension to the C programming language introduces new keywords and expressions that enable the user to distinguish between variables and functions that are stored in RAM and execute on the CPU (the host), and variables and functions (typically called kernels) that are stored and executed on the graphics processing unit (the device). Likewise, special keywords and expressions are available to transfer data between the host and the device.

The graphics processing unit that supports CUDA is composed of streaming multiprocessor, each of which is further composed of several scalar processors. Each streaming multiprocessor has different memory types available to it, namely a set of 32-bit registers, a shared memory block, as well as global memory. The 32-bit registers are fastest and smallest, while the global memory is largest and slowest. A scalar processor can access only its own register, each scalar processor in a particular streaming multiprocessor can access its own shared memory block, while the global memory is accessible to all scalar processors in all streaming multiprocessor (hence the whole device) as well as to the host. An important part of efficient parallelization of the problem involves minimizing the need to access global memory, and to make full use of the registers and the shared memory block. Although respecting the memory hierarchy is key to a fully optimized solution, significant improvements in computing power are attainable even if the memory is handled less than optimally.

From the programming point of view, each problem needs to be split into parts that can then be processed in parallel by threads. Threads form blocks of threads, and blocks further form grids of blocks. The total number of threads is thus the number of threads in each block times the numbers of blocks in the grid. Each thread is executed by a scalar processor, and each block of threads is assigned to a particular streaming multiprocessor. All threads within a block can thus access the previously mentioned shared memory block. The code that is executed within each thread is called a kernel, and each thread has a unique ID that is determined based on the block and grid structure. These are the basic concepts of parallel computing that quickly become familiar to those that decide to implement it.

\section{The spatial public goods game}

\begin{figure}
\centering{\includegraphics[width = 5.8cm]{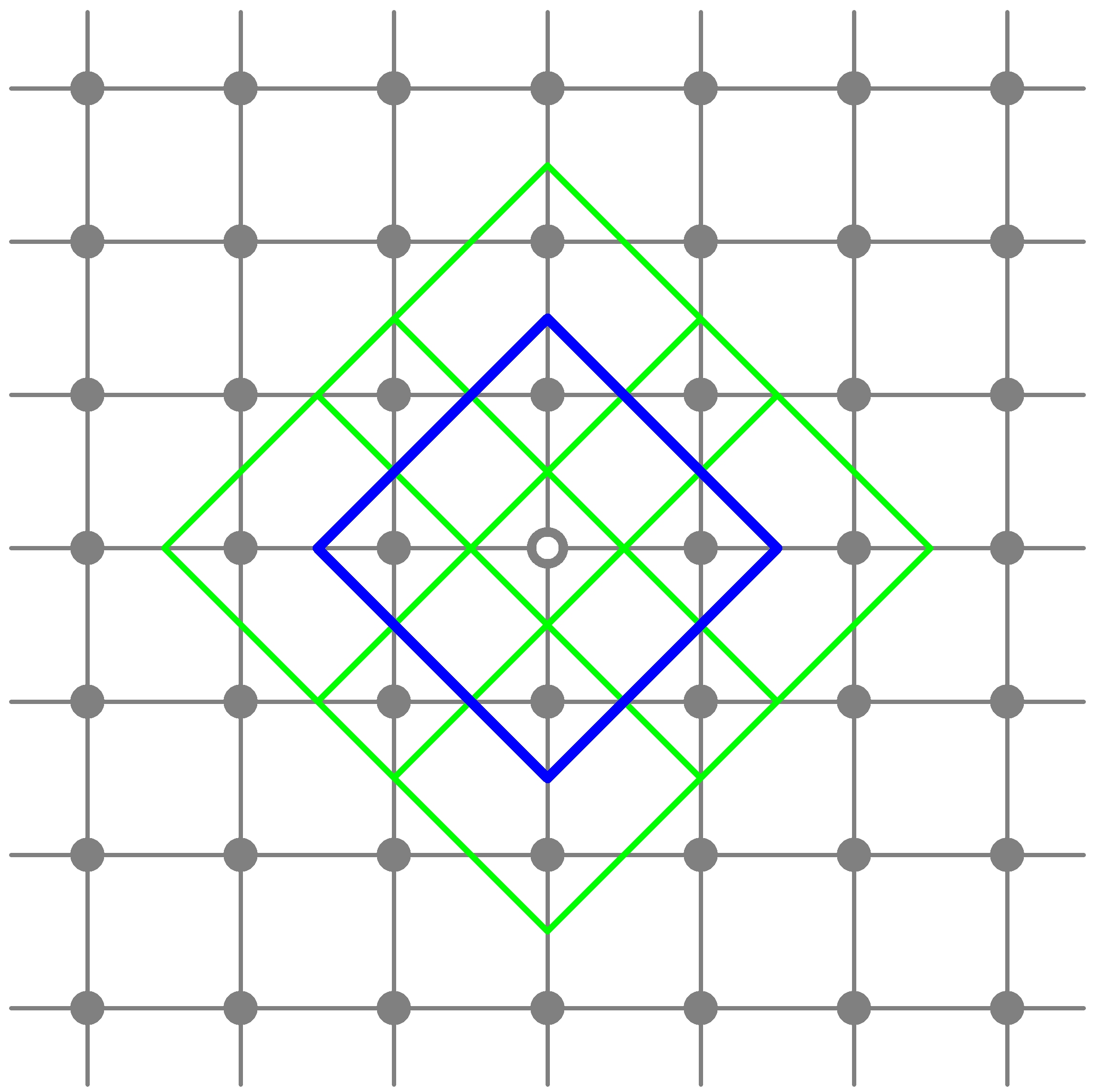}}
\caption{Schematic illustration of the overlapping groups of size $G=5$ around the player $s_x$, which is denoted with an open circle in the middle of the figure. The group where player $s_x$ is central and surrounded by its four nearest neighbors on the square lattice is marked with a thick blue line. The same player $s_x$ is also member in four other groups, which are denoted by thinner green lines. Due to the overlap of the groups the green lines denoting each particular group overlap as well. At a particular instance of the public goods game, the player $s_x$ obtains a payoff $\Pi_{s_x}^{g}$ from each of the depicted $g=1, \ldots, G$ groups. The overall payoff of player $s_x$ obtained at a particular instance of the game is thus $\Pi_{s_x}=\sum_g \Pi_{s_x}^{g}$.}
\label{schema}
\end{figure}

We use the spatial public goods game \cite{szabo_prl02, szolnoki_pre09c, perc_jrsi13} to demonstrate the parallelization of the Monte Carlo algorithm within the CUDA programming environment. The game is staged on a $L \times L$ square lattice with periodic boundary conditions where $L^2$ players are arranged into overlapping groups of size $G=5$, such that everyone is connected to its four nearest neighbors. As schematically illustrated in Fig.~\ref{schema}, each player $x=1, \ldots, L^2$ is therefore member in $g=1, \ldots, G$ different groups. Players that cooperate ($s_x=C$) contribute $c=1$ into the common pool, while players that defect ($s_x=D$) contribute nothing. The sum of all contributions within a group is multiplied by $r>1$ and then divided equally amongst all group members regardless of their strategies. In a group $g$ containing $G$ players, of which $N_C$ cooperate, the resulting payoffs for cooperators and defectors are thus
\begin{eqnarray}
\Pi_C^{g} &=& G^{-1}rcN_C-c\,\,\,\,{\rm and}\\
\Pi_D^{g} &=& G^{-1}rcN_C,
\end{eqnarray}
respectively. Evidently, the payoff of a defector is always larger than the payoff of a cooperator, if only $r<G$. With a single parameter, the public goods game hence captures the essence of a social dilemma in that defection yields highest short-term individual payoffs, while cooperation is optimal for the group, and in fact for the society as a whole. The overall payoff $\Pi_{s_x}$ of a player $x$ from all the $g=1, \ldots, G$ groups is simply the sum $\Pi_{s_x}=\sum_g \Pi_{s_x}^{g}$.

Monte Carlo simulations of the described public goods game are carried out as follows. Initially each player on site $x$ of the $L \times L$ square lattice is designated either as a cooperator ($s_x = C$) or defector ($s_x = D$) with equal probability. The following elementary steps are subsequently repeated in a random sequential manner. A randomly selected player $x$ plays the public goods game as a member of all the $g=1,\ldots,G$ groups, thereby obtaining the payoff $\Pi_{s_x}$. Next, one of the four nearest neighbors of player $x$ is chosen uniformly at random, and this player $y$ acquires its payoff $\Pi_{s_y}$ in the same way. Finally, player $x$ copies the strategy $s_y$ of its randomly chosen nearest neighbor with the probability determined by the Fermi function
\begin{equation}
W(s_y \to s_x)=\frac{1}{1+\exp[(\Pi_{s_x}-\Pi_{s_y})/K]},
\end{equation}
where $K$ quantifies the uncertainty by strategy adoptions \cite{szabo_pre98, szolnoki_pre09c}. In the $K \to 0$ limit, player $x$ copies the strategy of player $y$ if and only if $\Pi_{s_y} > \Pi_{s_x}$. Conversely, in the $K \to \infty$ limit, payoffs seize to matter and strategies change as per flip of a coin. Between these two extremes players with a higher payoff will be readily imitated, although the strategy of under-performing players may also be occasionally adopted, for example due to errors in the decision making, imperfect information, and external influences that may adversely affect the evaluation of an opponent. Repeating all the described elementary steps $L^2$ times constitutes one full Monte Carlo step (MCS), thus giving a chance to every player to change its strategy once on average.

As the main observable, we determine the average fraction of cooperators as
\begin{equation}
\rho_C=<\frac{1}{L^2}\sum_{x=1}^{L^2}d_x>,
\end{equation}
where $d_x=1$ if $s_x=C$ and $d_x=0$ otherwise, and $<\ldots>$ indicates average over time in the stationary state. A sufficiently long relaxation time needs to be discarded prior to this. In general, the stationary state is reached once $\rho_C$ becomes independent of the time interval over which it is determined.

\section{Parallelization of the Monte Carlo simulation method}

\begin{figure}
\centering{\includegraphics[width = 8.5cm]{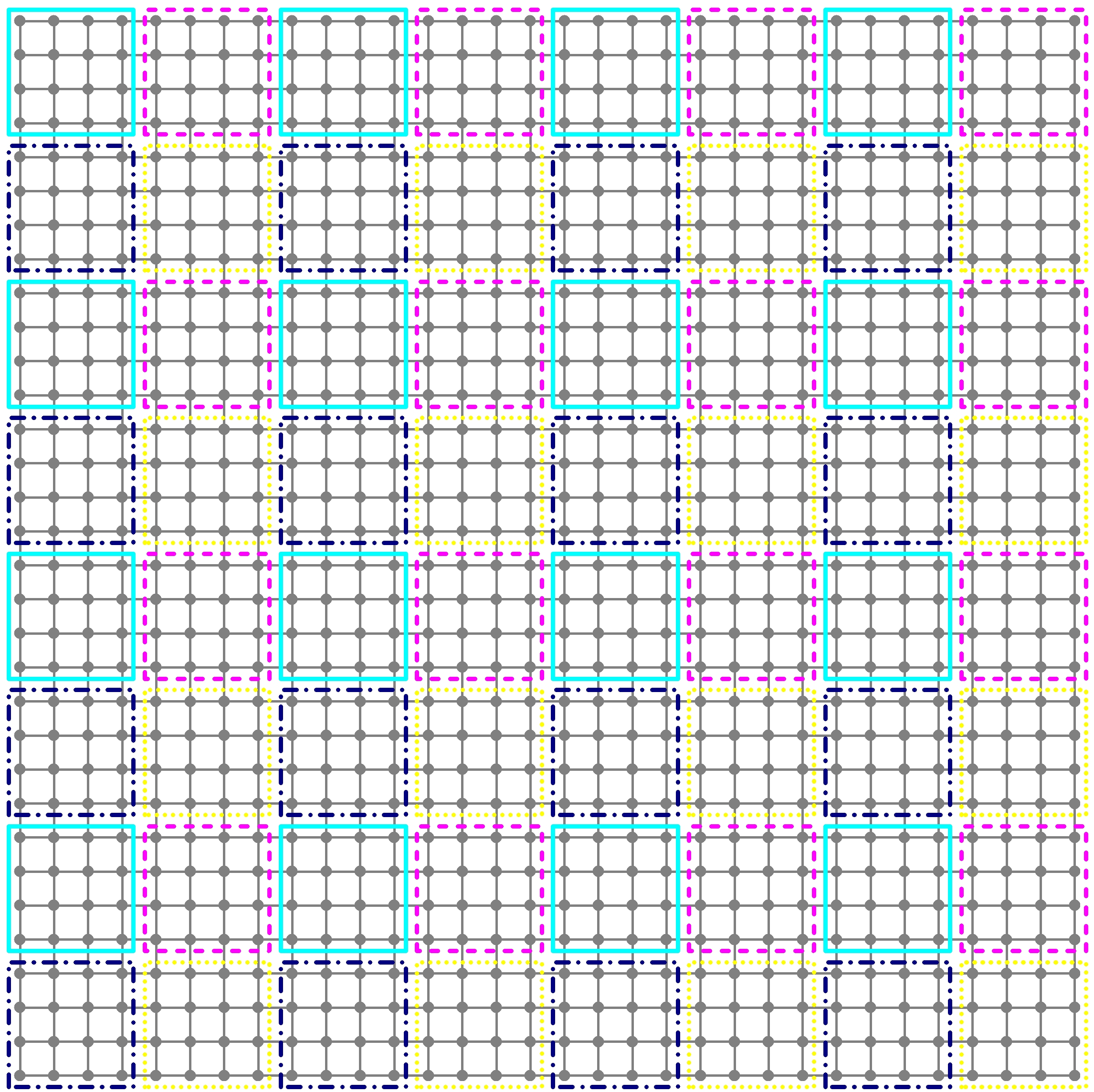}}
\caption{Schematic illustration of the double tiling decomposition scheme on a $32 \times 32$ square lattice, where each domain comprises $4 \times 4$ players. These equally sized and independent domains are marked with squares of different line style and color, such that $\{1,2,3,4\}=\{C,M,Y,K\}$. Each full Monte Carlo step is effectively split into four parts, the first part updating players within all cyan domains, the second part updating players within all magenta domains, the third part updating players within all yellow domains, and finally the fourth part updating players within all key domains. Importantly, for this decomposition scheme to work out, the linear size of the whole lattice $L$ needs to be exactly divisible by two times the linear size of each domain ($8$ in this case). There are exactly $T=L^2/4$ domains of a particular color on the whole lattice, and the strategies of the players within these domains can be updated in parallel by $T$ threads. How the number of these threads is distributed within blocks, and how many such blocks are needed to form the grid of the graphics processing unit depends on the architecture of the graphics card. We have chosen each block to contain $32 \times 2$ threads, which is related to the warp size on current graphics cards and the size of the shared memory block per streaming multiprocess.}
\label{partition}
\end{figure}

While the implementation of the Monte Carlo simulation method described in Section~III is straightforward, its parallelization requires care in that we have to make certain that threads that will process different parts of the lattice do not simultaneously change strategies of the same players, and that the strategies of players do not change whilst the determination of the payoffs takes place. The remedy lies in partitioning the lattice as shown in Fig.~\ref{partition}, which was originally proposed in \cite{shim2005semirigorous} for parallel kinetic Monte Carlo simulations of thin film growth, and subsequently used also in \cite{kelling2012comparison} for simulating the Kardar-Parisi-Zhang model and the kinetic Monte Carlo model. The approach is known as the double tiling decomposition scheme, effectively partitioning the lattice into $T=L^2/4$ equally sized and independent domains of type $1$, $2$, $3$ and $4$ (see Fig.~\ref{partition}), which during a full Monte Carlo step should be updated in turn. Accordingly, $T=L^2/4$ gives us the number of threads needed in total within the graphics processing unit, with each thread being assigned to one domain. Note that a full Monte Carlo step is split into four parts, the first part updating players within all domains $1$, the second part updating players within all domains $2$, the third part updating players within all domains $3$, and finally the fourth part updating players within all domains $4$. But since all the domains of a given type are independent in that they do not share a player or even a border, they can be updated in parallel by $T=L^2/4$ threads.

\begin{figure}
\centering{\includegraphics[width = 5.8cm]{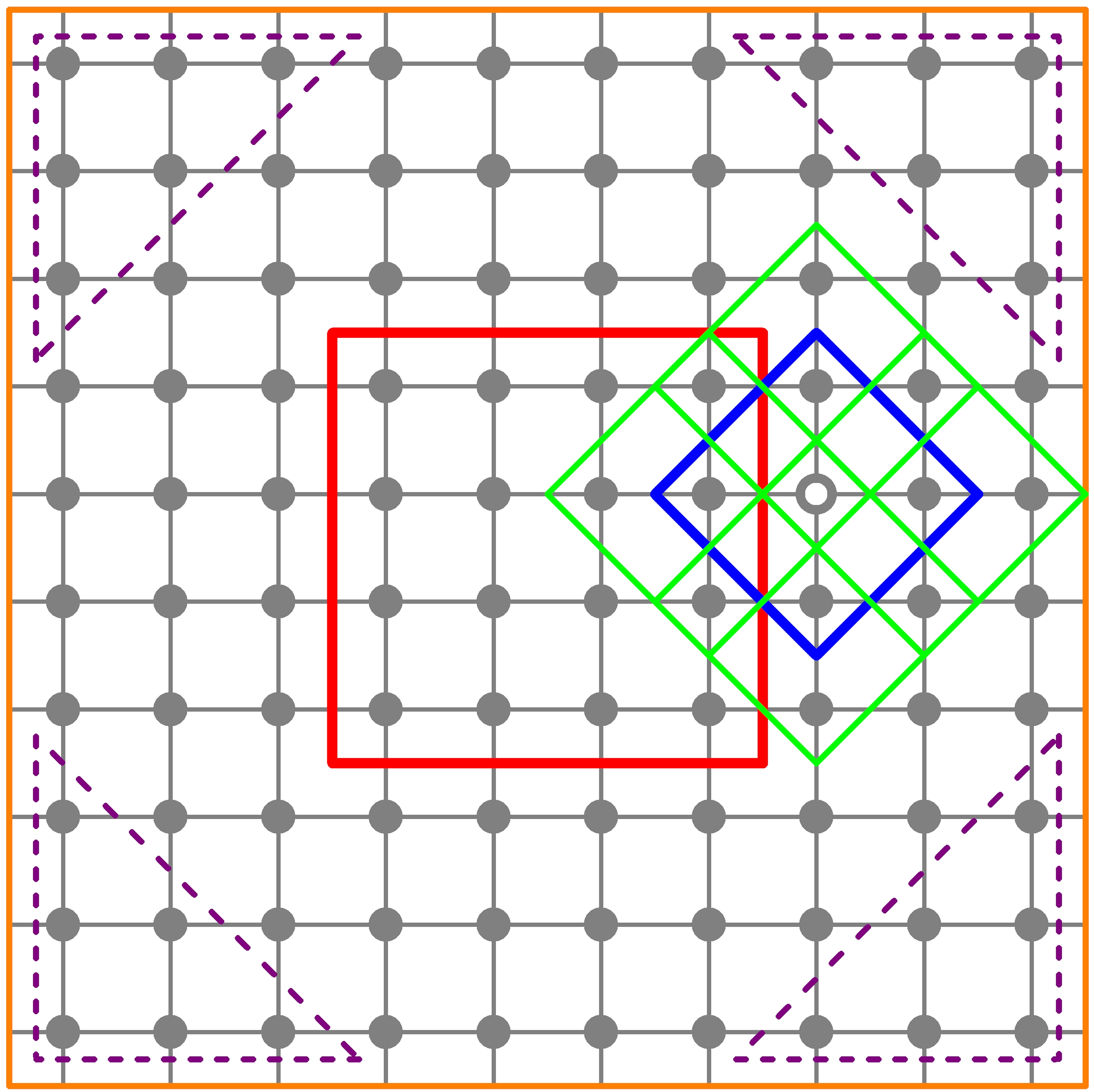}}
\caption{Schematic illustration of the memory that needs to be available to a thread for the strategies of players within the $4 \times 4$ red domain to be updated correctly. The memory block needed is marked with a thinner orange line. If the strategies of all the players within the orange domain are available to the thread, then whichever player $x$ within the $4 \times 4$ red domain is randomly selected to potentially copy the strategy from one of its randomly chosen nearest neighbors $y$, this ensures that both payoffs $\Pi_{s_x}$ and $\Pi_{s_y}$ entering the Fermi function given by Eq.~(3) are determined in full. To be precise, the strategies of the players encircled with dashed violet triangles in each corner of the orange domain are actually not needed, which can be verified if the overlapping groups schematically illustrated in Fig.~\ref{schema} are superimposed on $y$ players laying outside of the $4 \times 4$ red domain (one such example is shown for clarity). For the simplicity and efficiency of the source code, it is however not worthy of implementation to filter these players out of the memory block.}
\label{memory}
\end{figure}

However, by looking at the schematic display of groups in Fig.~\ref{schema}, it becomes clear that a randomly selected player $x$ at the border of a particular domain will need some information of player strategies also from adjacent domains. Even more so if the potential source of the new strategy, player $y$, will be selected on the other side of the border. We remind that player $y$ can be any of the four nearest neighbors, and a player $x$ at the border of the domain will have one player as the nearest neighbor in the other domain (players in the corners will in fact have two nearest neighbor in two other domain). The memory that needs to be passed to a particular thread must thus contain not only the strategies within a domain, but also the strategies of players three lines outward in each direction. This is schematically illustrated in Fig.~\ref{memory}. The reader can convince herself that this is in fact the case by following the composition of groups depicted in Fig.~\ref{schema} along the border of a particular domain whilst assuming that player $y$ is chosen from the other side of the border. Thankfully, this requirement does not interfere with independent parallel random sequential updating in the other domains of the same color in Fig.~\ref{partition} if only the size of the domains is $3 \times 3$ or larger.

This is essentially all there is to the parallelization. Technically, we thus split the lattice into $T=L^2/4$ adjacent domains of four different types, update all domains of the same type in parallel, and do so consecutively over all the type to complete one full Monte Carlo step. The only technical hiccup is to make sure information on player strategies is available also three player lines outward in each direction. In our particular case, we have chosen the domain size $S$ to be $4 \times 4$ (exactly as displayed in Fig.~\ref{memory}), which together with the auxiliary memory still allows all strategy to be stored in the fast shared memory block if the number of threads within a block $B$ is $32 \times 2$. The latter choice, on the other hand, is conditioned on the fact that one streaming multiprocessor simultaneously executes a so-called warp of $32$ threads in the large majority of today's graphic cards. Given these choices, the number of blocks within the grid can be determined according to $G=T/(BS)$. We note that, depending on $L$, $G$ might not be an integer. In fact, it will be only when $L$ is exactly divisible by $64$. In all other cases $G$ should simply be the smallest integral value not less than the originally calculated $G$. In this case one ends up with more threads then needed, but it is easy to discard those with an $if$ statement. For example, for $L=800$ and the domain size $S=16$, it is clear that the actual number of threads needed is $L^2/4S=10^4$. But with the ``number of threads within a block'' $B=64$ constrain, the number of blocks within the grid will be $G=157$, in turn yielding $157B=10048>10^4$ threads available.

Lastly, we note that the selection of the $4 \times 4$ domain size and the partitioning of the lattice linearly in sequences of $2$ ($1,2,1,2,1,\ldots$ or $3,4,3,4,3,\ldots$) obviously requires that the linear size of the lattice $L$ be divisible by $8$ lest the decomposition will not be perfect. We again refer to Fig.~\ref{partition} for details.

The source code that implements the above described parallelization is available as supplementary material to this paper as well as at \href{http://github.com/matjazperc/pgg}{\textcolor{blue}{github.com/matjazperc/pgg}}.

\section{Results}
As explained in Section~III when introducing the payoffs of the public goods game, if $r<G$ the payoff of a defector is always larger than the payoff of a cooperator. Accordingly, $r=G$ is the threshold that marks the transition between defection and cooperation in well-mixed populations, where groups are formed by selecting players uniformly at random. In structured populations, however, due to the so-called network reciprocity \cite{nowak_s06}, cooperators are able to survive at multiplication factors that are well below the $r=G$ limit that applies to well-mixed populations. The manifestation of network reciprocity relies on pattern formation, such that cooperators form compact clusters and can thus avoid exploitation by defectors \cite{nowak_n92b}. In short, cooperators do better if they are surrounded by other cooperators.

\begin{figure}
\centering{\includegraphics[width = 8.5cm]{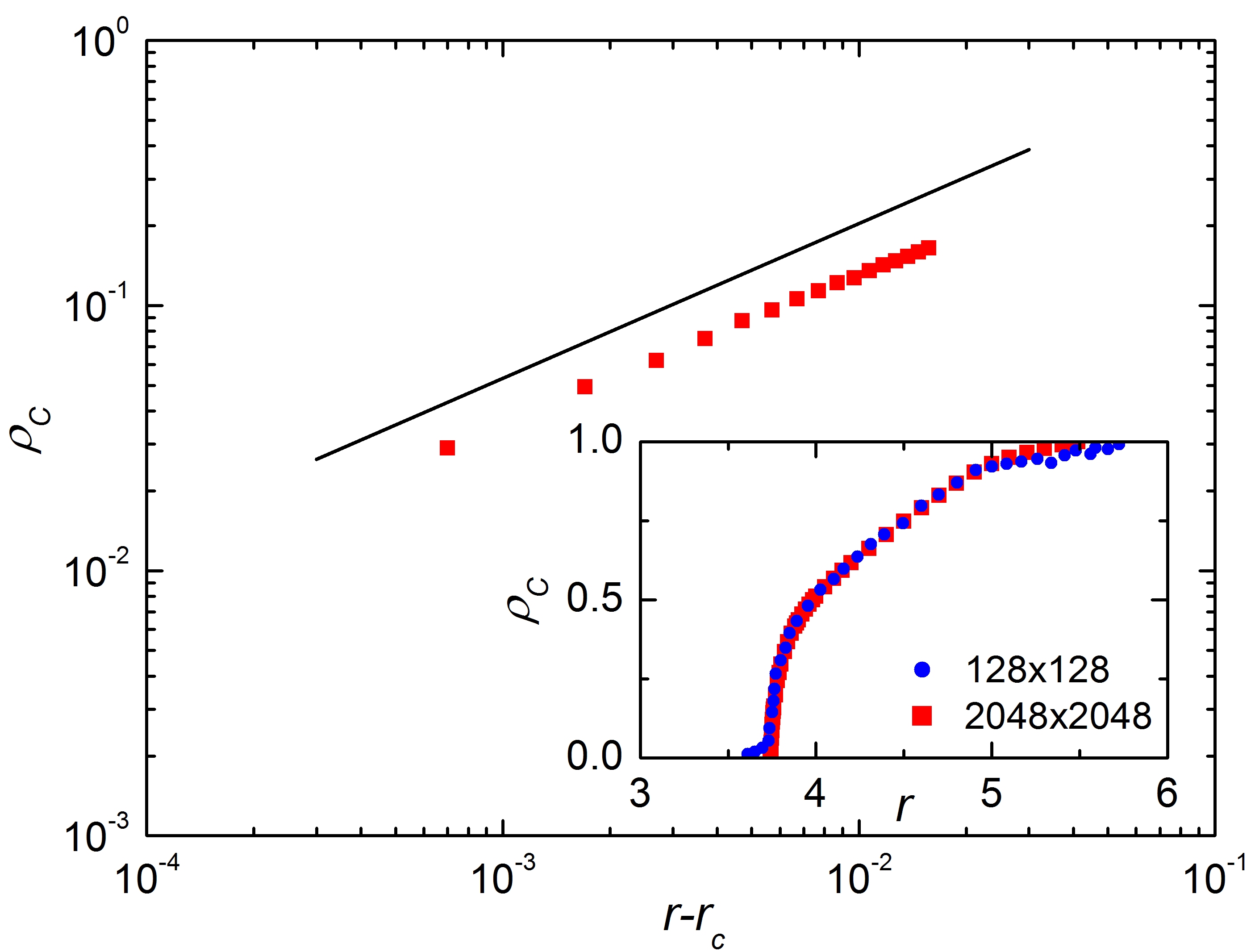}}
\caption{Second-order $C+D \to D$ phase transition and the confirmation of the directed percolation universality class conjecture in the public goods game on the square lattice. The main panel shows the fraction of cooperators $\rho_C$ in the stationary state in dependence on the distance to the critical value of the multiplication factor $r_c=3.7443(4)$ in log-log scale. The black line indicates the slope $0.584$ characteristic for directed percolation, while a linear fit to the data yields $\beta=0.56(2)$. The inset shows the fraction of cooperators $\rho_C$ in the stationary state in dependence on the value of the multiplication factor $r$. It can be observed that the $C+D \to D$ phase transition from right to left as $r$ decreases is continuous. Blue circles show that the nature of both phase transitions is distorted if staying at a small $128 \times 128$ size across the whole span of $r$ values, thus corroborating the need for larger lattices near phase transition points.}
\label{critical}
\end{figure}

Previous research has shown that for the spatial public goods game on the square lattice with $K=0.5$ the threshold for cooperators to survive is $r>3.74$ \cite{szolnoki_pre09c}. Also importantly, it was shown that the public goods game on the square lattice exhibits continuous phase transitions that belong to the directed percolation universality class \cite{helbing_njp10}, such that
\begin{equation}
\rho_{s_x} \propto |p-p_c|^\beta
\end{equation}
where $p_c$ is the critical parameter value at which the absorbing phase is reached and $\beta=0.584(4)$ is the critical exponent \cite{hinrichsen_ap00}. We remind the reader that the directed percolation universality class conjecture requires that \cite{janssen1981nonequilibrium, grassberger1982phase} (i) the model displays a continuous phase transition from a fluctuating active phase into a unique absorbing phase, (ii) the transition is characterized by a positive one-component order parameter (in our case $\rho_C$), (iii) the dynamic rules involve only short-range processes (yes due to the square lattice), and (iv) the system has no special attributes such as additional symmetries or quenched randomness.

We can use this conjecture that fully applies to the presently studied public goods game as validation for the parallelization of the Monte Carlo method presented in Section~IV. As shown in Fig.~\ref{critical}, the phase transition leading from the mixed $C+D$ phase at $r>3.74$ to the absorbing $D$ phase as $r$ decreases is indeed continuous (see inset), and it belongs to the directed percolation universality class with $r_c=3.7443(4)$ and $\beta=0.56(2)$ (main panel). This confirms the previously published results in \cite{szolnoki_pre09c, helbing_njp10} obtained with traditional CPU-based computing, and it also confirms the applicability of the directed percolation universality class conjecture to the public goods game on the square lattice. For the results presented in Fig.~\ref{critical}, we have simulated the public goods game on lattices of size $2048 \times 2048$ in the immediate  proximity of the phase transition point up to $3 \times 10^6$ full MCS, subsequently using smaller lattice sizes and shorter simulations times when being further away from $r_c$. With CPU-based computing, such simulations typically require weeks to complete on clusters with many cores. As the inset of Fig.~\ref{critical} shows, without using larger lattices, for example staying at the $128 \times 128$ size where CPU-based computing is not significantly slower than parallel computing, the nature of the phase transition becomes distorted. We refer to page $251$ in \cite{krauth2006statistical} for a classical statistical mechanics example of qualitatively the same phenomenon.

\begin{figure}
\centering{\includegraphics[width = 8.5cm]{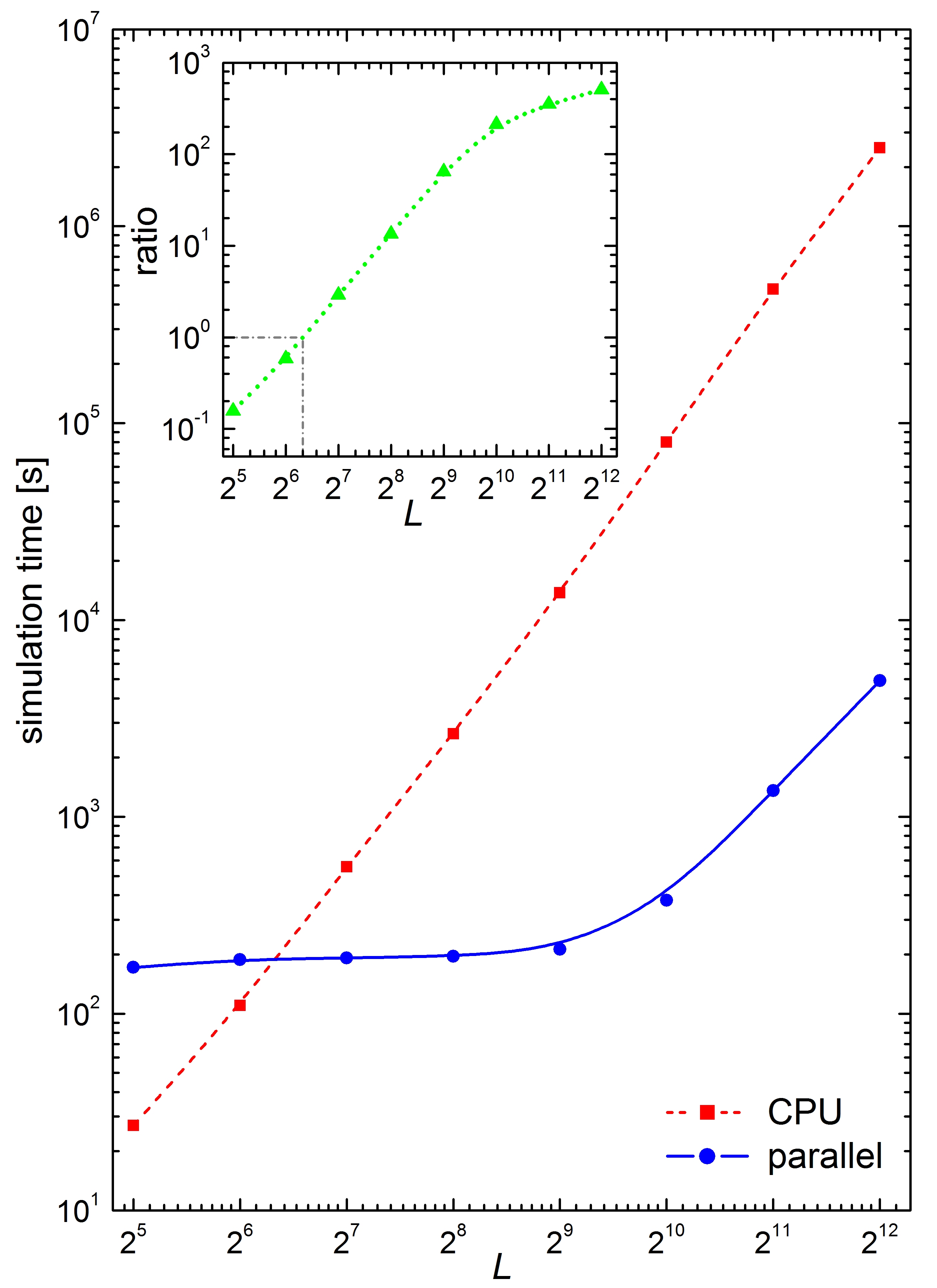}}
\caption{Quantitative comparison of Monte Carlo simulation times between CPU-based computing and parallel computing (see legend). As the benchmark, we have simulated the public goods game on the $L \times L$ square lattice for $10^6$ full MCS, using $K=0.5$ and $r=3.98$ at which the fraction of cooperators in the stationary state is $\rho_C=0.5$. CPU-based computing was done using an Intel Xeon processor with 2.80~GHz clock speed, while parallel computing was done using NVIDIA GeForce GTX 1080 graphic card with $2560$ streaming multiprocessor each running at 1607~Mhz closk speed. The main panel shows the increase in the length of the simulation time for CPU-based computing (red squares) and parallel computing (blue circles) while going from $L=32$ to $L=4096$. The inset shows the ratio between parallel computing time and CPU-based computing time for the same span of $L$ values. The dashed-dotted gray line in the inset marks the lattice size $L_e=80$ at which CPU-based computing is equally fast as parallel computing. Below $L_e$ CPU-based computing is faster, while above this lattice size parallel computing is faster. Lines are just to guide the eye.}
\label{accelerate}
\end{figure}

It remains of interest to quantitatively asses the advantage in time obtained by switching from CPU-based computing to parallel computing on the graphics processing unit. Results presented in Fig.~\ref{accelerate} show that for CPU-based computing the simulation times increase roughly with the square of the increase of the linear size of the lattice $L$. For example, if using the same number of full MCS, it takes approximately four times as much time to finish the simulation on a $256 \times 256$ square lattice as it does to finish the same simulation on the $128 \times 128$ square lattice. As the chosen linear size of the lattice grows, this factor of four grows ever so slightly towards five and above due to increasing memory demands. Note that at $2048 \times 2048$ lattice size it takes in excess of five days for the CPU to complete $10^6$ full MCS. At $4096 \times 4096$ lattice size it takes nearly a month. Conversely, with parallel computing the simulation times remain nearly constant as $L$ increases to the point where the number of threads does not exceed $2560$ streaming multiprocessors present in the graphics card that we have used (NVIDIA GeForce GTX 1080). With the $4 \times 4$ domain size the double tiling decomposition scheme, this yields a $L=400$, beyond which the simulation times start increasing with a factor between three and five for each doubling of the linear lattice size $L$. Accordingly, the larger the lattice size the greater the benefits from parallel computing.

This is confirmed in the inset of Fig.~\ref{accelerate}, where the ratio between parallel computing time and CPU-based computing time in dependence on $L$ is shown. We note that the ratio can also be smaller than one, in particular if the lattice is so small that the benefits of parallel computing can not be taken advantage of with the constrain of a $4 \times 4$ domain size (we remind the reader that $3 \times 3$ is the smallest permissible domain size to avoid prohibited memory overlaps between the domains in Fig.~\ref{partition}, as explained with Fig.~\ref{memory}). In that case the number of threads does not offset the slower clock speed of streaming multiprocessors in comparison to the CPU clock speed (Intel Xeon processor with 2.80~GHz clock speed). The dashed-dotted gray line in the inset at $L_e=80$ marks the lattice size at which CPU-based computing is equally fast as parallel computing. When $L>L_e$, however, the acceleration of the simulations is impressive. At $L=4096$, the ratio is $\approx 500$, thus cutting the simulation time for $10^6$ full MCS from a month to less than an hour and a half at this lattice size. Of course still somewhat too long for classroom demonstrations, but such large lattice sizes are rarely needed (see \cite{szolnoki_prx13} for an example). Taken together, results presented in Fig.~\ref{accelerate} confirm that the effort in successfully parallelizing the Monte Carlo simulation method is rewarded with simulation times that can be orders of magnitude shorter than attainable with conventional CPU-based computing.

\section{Discussion}
We have advocated for the use of graphics processing units to bring high-performance parallel computing into the physics classroom. We have used the public goods game on the square lattice as an example to demonstrate the parallelization of the Monte Carlo simulation method by means of the double tiling decomposition scheme, and we have explained in detail the subtleties of memory sharing and conflict avoidances when simultaneously updating different parts of the lattice by several threads running in parallel at the same time. We have shown that the parallelization preserves all the most important properties of the public goods game related to statistical physics, in particular the continuous character of the $C+D \to D$ phase transition and the directed percolation universality class to which the phase transition belongs. While the calculation of these results would require weeks of many-core clusters if done with traditional CPU-based computing, a single capable graphics card decreases this time by a factor of 500, thus making these phenomena viable for presentation in the classroom.

We note that the described parallelization scheme can be easily adapted so that Monte Carlo simulations of other evolutionary games on the square lattice can be performed, such as for example the well-known prisoner's dilemma game \cite{hauert_ajp05, javarone2016statistical} or other social dilemmas \cite{szabo_pr07, perc_bs10}. An important difference with regards to the public goods game is that the prisoner's dilemma game is played in a pairwise manner, not in groups. As a consequence, the memory needed to update strategies within a given domain is one line outward less in each direction (see Fig.~\ref{memory}), which can serve as good practice when adapting the source code. Of course, and as already demonstrated in original research published in the past \cite{preis2009gpu, janus2010acc, schulz2011simulation, kelling2011extremely, seibert2011mapping, barros2011simulation, kelling2012comparison, juhasz2012anomalous, polyakov2013large, odor2014aging, spiechowicz2015gpu}, the same approach can be used to simulate a broad variety of classical statistical physics systems.

With the Moore's law, stating that the number of transistors in a dense integrated circuit doubles approximately every two years, arriving at it limits due to fundamental technical constraints in CPU design, today's hardware has to be designed in a multi-core manner to keep up. This in turn means that if we want to benefit from faster simulation times, the software has to be written in a multi-threaded manner to take full advantage of the hardware. The graphic cards industry has invested admirable effort for this to materialize, one product of which is the friendly and thoroughly documented CUDA programming environment. The time is thus ripe for these programming techniques to be introduced into graduate and advanced undergraduate curricula, giving students the chance to learn the benefits of parallel computing from the onset of their physics education. We hope that this paper will be useful to that effect.

\end{document}